# Luminosity indicators in dusty photoionized environments

Mark Bottorff, Joseph LaMothe, Emmanuel Momjian, Ekaterina Verner, Dejan Vinkovic', & Gary Ferland

Physics Department, University of Kentucky
Lexington, KY 40506

Abstract

The luminosity of the central source in ionizing radiation is an essential parameter in a photoionized environment, and one of the most fundamental physical quantities one can measure. We outline a method of determining luminosity for *any* emission-line region using only infrared data. In dusty environments, grains compete with hydrogen in absorbing continuum radiation. Grains produce infrared emission, and hydrogen produces recombination lines. We have computed a very large variety of photoionization models, using ranges of abundances, grain mixtures, ionizing continua, densities, and ionization parameters. The conditions were appropriate for such diverse objects as HII regions, planetary nebulae, starburst galaxies, and the narrow and broad line regions of active nuclei. The ratio of the total thermal grain emission relative to H$\beta$ (IR/H$\beta$) is the primary indicator of whether the cloud behaves as a classical Strömgren sphere (a hydrogen-bounded nebula) or whether grains absorb most of the incident continuum (a dust-bounded nebula). We find two global limits: when IR/H$\beta$ < 100 infrared recombination lines determine the source luminosity in ionizing photons; when IR/H$\beta \gg$ 100 the grains act as a bolometer to measure the luminosity.

Subject Headings: Infrared: General, Infrared: Stars, ISM: HII Regions, ISM: Planetary Nebulae, Methods: Observational, Stars: Fundamental Parameters

# 1. Introduction

With the development of modern infrared detectors, it is now common to detect photoionized regions where dust is not only clearly present, but also prevents visible light from escaping. Examples include starburst galaxies, the central regions of nearly all galaxies (Engelbracht, Reicke, & Reicke 1997), and obscured galactic nebulae (Van de Sleene & Pottash 1995). For heavily obscured regions infrared observations are often the only source of information. In these cases it is necessary to develop new methods of determining fundamental parameters such as the mass of ionized gas or the luminosity of the continuum source. First steps in this direction were taken by Spinoglio & Malkan (1992) and Voit (1992), who looked at potential implications of upcoming ISO data.

Grains effect photoionized nebulae in several ways. 1) They compete with hydrogen in directly absorbing the incident continuum. 2) Energetic photons photoionize the grains and the photoelectrons heat the gas. 3) They reradiate the absorbed radiation, producing a characteristic thermal continuum. Oliveira & Maciel (1986), Borkowski & Harrington (1991), and Baldwin et al. (1991) discuss various aspects of grain heating, cooling, and emission. Because of these effects, especially the direct absorption of the continuum, standard methods of determining the continuum luminosity (Osterbrock 1989) may not be valid in dusty environments.

The goal of this paper is to determine diagnostic indicators that will make it possible to measure the luminosity in ionizing radiation from infrared data alone. We present results of a very large number of photoionization models to study the various affects grains can have. These models cover very wide ranges of density, ionization, composition, and sources of ionization. Our calculations include conditions found in HII regions, planetary nebulae, starburst galaxies, and the narrow and broad line regions of active nuclei. We find two global correlations, which together provide a method to measure the ionizing continuum luminosity to within a factor of two.

## 2. Photoionization calculations

Photoionized nebulae are non-equilibrium plasmas, and their spectra are best studied through full numerical simulation. Photoionization models are by necessity idealized; the incident continuum must be assumed, chemical and grain homogeneity are often taken for granted, and the geometry must be specified. A real nebula is the result of the history of events that occurred over its lifetime and may be far more complicated than we imagine. Our approach is to consider the widest possible variety of photoionization models, and then look for global characteristics. We have identified correlations that occur despite the breadth of models that are considered. We can therefore be relatively confident that many situations occurring in nature are included in our calculations. A similar approach was used to determine helium ionization correction factors, a quantity needed for measurements of accurate nebular helium abundances (Ali et al. 1991).

## 2.1. Expectations from simple theory

In a grain-free environment, hydrogen provides the dominant source of opacity. It absorbs nearly all of the radiation shortward of 912Å. In this case the simple photon counting arguments given below provide a direct method of counting the number of ionizing photons (Osterbrock 1989).

For simplicity consider a plane-parallel geometry with a flux of hydrogen ionizing photons $\phi(H)$ [cm$^{-2}$ s$^{-1}$] striking the illuminated face of the cloud. The hydrogen ionization front occurs at the Strömgren depth L. In a grain-free environment the depth is set by the balance between the rate ionizing photons enter the cloud and the number of recombinations that occur over L:

$$\phi(H) = n_e n_p \alpha_B(T) L \ . \tag{1}$$

Here $n_e$ and $n_p$ are the electron and proton densities, and $\alpha_B$ is the case B hydrogen recombination coefficient (Storey & Hummer 1995). In this case the Strömgren depth L increases linearly with $\phi(H)$. A hydrogen recombination line's total emission is given by

$$4\pi J_{u,l} = n_e n_p \left( \frac{4\pi j_{u,l}(T)}{n_e n_p} \right) L = \phi(H) \left( \frac{4\pi j_{u,l}(T)}{n_e n_p} \right) \Big/ \alpha_B(T) \ \ [\text{erg cm}^{-2}\ \text{s}^{-1}]. \tag{2}$$

The term in parenthesis, $4\pi j_{u,l}/n_e n_p$, can be obtained for any hydrogen line (Storey & Hummer 1995). In the pure hydrogen limit given by equation 2 the intensity of a hydrogen recombination line is directly proportional to the flux in ionizing photons, so the line luminosity probes the flux of ionizing photons. The luminosity of the continuum source can be then estimated if the continuum shape is even roughly known.

In the rest of this paper we will present our results relative to the optical Hβ line, since this is standard in numerical simulations of nebulae. This line would not be used in practice since the optical extinction could be quite large, especially if intervening material is present. Studies of dusty HII regions show that the intrinsic hydrogen recombination spectrum remains close to Case B predictions (Hummer & Storey 1992). Our results could be rescaled to any other hydrogen line using standard tables (Storey & Hummer 1995). For instance, the intrinsic intensities of Pα and Brγ are 0.34 and 0.028 times the intensity of Hβ.

Only the quantity $4\pi J_{u,l}$ is directly observed, and $\phi(H)$ is inferred from it. Equation 2 underestimates $\phi(H)$ in a dusty cloud since grains, rather than hydrogen, absorb the incident continuum. The last substitution in equation 2 uses equation 1 and is incorrect when grains absorb the incident continuum.

Then, only a fraction of the total $\phi(H)$ is absorbed by hydrogen. This reduces the Strömgren length $L$ and makes the lines less intense for a given $\phi(H)$. The continuum luminosity will be underestimated.

Figure 1 shows two of the grain absorption opacity functions we use below. As is standard in the interstellar medium literature they are expressed as the opacity (cm$^{-1}$) per unit volume of hydrogen. The solid line represents the Orion grain mixture described by Baldwin et al. (1991), and the dashed line is an opacity function deduced by Kevin Volk for young planetary nebulae (PNe; Volk 1991). The optical depth due to grains at any particular wavelength is the function shown in Figure 1 multiplied by the total hydrogen column density. The effects of grains depend mainly on their optical depth across the Strömgren thickness. Grains should have little effect if this optical depth is much less than one. The actual column density needed before grains absorb a significant fraction of the incident continuum will depend somewhat on the continuum shape since the opacity depends on wavelength. Ionized hydrogen column densities of $N_H = 10^{21}$ cm$^{-2}$ or greater should be thick enough for grains to absorb some light, and results in an underestimate of the continuum luminosity.

From the above discussion, it is clear that the total column density through the ionized layer is a basic concern. Equation 1 can be rewritten in terms of the hydrogen column density

$$N_H = n_p L = \frac{\phi(H)}{n_e \alpha_B(T)} = U \frac{c}{\alpha_B(T)} \approx U \, 10^{23} \; [\text{cm}^{-2}] \tag{3}$$

where we have introduced the dimensionless ionization parameter $U$, the ratio of photon to electron densities,

$$U = \frac{\phi(H)}{n_e c} \; . \tag{4}$$

The result is that the larger ionization parameter models have larger ionized column densities, larger grain column densities, and so the grains absorb a greater fraction of the incident continuum. For the grain opacity functions used, nebulae with ionization parameters above $U = 10^{-2}$ are increasingly dust-bounded. At the largest $U$ the continuum will be almost totally absorbed by grains and the standard Strömgren arguments (equation 2) become irrelevant. In this limit grains, not hydrogen, absorb the incident continuum in a fully dust-bounded nebula. The ionization parameter where grains begin to dominate depends on the dust to gas ratio, the grain opacity function and the shape of the incident ionizing continuum. This makes the results somewhat model dependent, so we need to explicitly explore a variety of models.

## 2.2. Photoionization models of an HII region

We use version 90.04a of the spectral synthesis code Cloudy, described by Ferland et al. (1998), and available on the web at http://www.pa.uky.edu/~gary/cloudy, in the calculations described below.

The arguments in the previous section show that grains and hydrogen compete in absorbing ionizing radiation, and that grains should be more important for larger ionization parameters. It is not obvious what trends will exist since the dust to gas ratio can vary from object to object, the form of the grain opacity function depends on the grain size distribution, and softer continua will be more efficiently absorbed since the grain opacity peaks near 912Å (Figure 1). We anticipate that the ratio of the total grain thermal emission to the intensity of a hydrogen emission should be a sharp function of the competition between grains vs hydrogen attenuating the continuum.

As an example we first look in detail at one of the many environments that we will explore below. We computed a large grid of models for conditions appropriate for parts of the Orion HII region. A later section will show that these results are general. The incident continuum is a Kurucz stellar atmosphere (Kurucz 1991), and the effective temperature is varied between 30,000 K and 50,000 K. The ionization parameter was varied over a range far larger than encountered in Orion itself: $-4 \leq \log(U) \leq 0.5$. The hydrogen density was kept constant at $n_H = 10^2$ cm$^{-3}$. The Orion abundances and grains described by Baldwin et al. (1991) and Ferland et al. (1998) were used. The abundances are roughly 2/3 of solar, with large depletions for refractory elements. The grains have a large ratio of total to selective extinction, representative of grains within the Orion environment. Radial integrations stopped when carbon became neutral. For the continuum sources considered here, this ensured that well over 99% of the incident radiation had been absorbed by gas or dust. The results are shown in Figure 2.

The top panel of Figure 2 shows that the ratio of the total thermal grain emission (designated simply IR) to the intensity of Hβ is a strong function of the ionization parameter. At low ionization parameters the nebulae are classical hydrogen Strömgren spheres, the grains absorb little of the incident continuum, and the IR/Hβ ratio is small. The ratio increases as $U$ increases and the nebula becomes increasingly dust-bounded. In absolute terms, as $U$ increases the IR continuum increases since grains absorb more radiation, and the Hβ intensity decreases since hydrogen gets fewer of the ionizing photons. The ratio shows a weaker dependence on the stellar continuum, with larger IR/Hβ at lower temperatures. These stars are cool enough for a significant fraction of their luminosity to occur at energies lower than 912Å. At constant $U$ the flux of photons shortward of 912Å is constant and the effect of decreasing the stellar temperature is to increase the radiation field longward of 912Å. This soft radiation heats the grains without ionizing hydrogen, thus increasing the ratio. The strongest dependence is on *U* however, and the IR/Hβ ratio will be the primary indicator of the role grains play in analysis that follows.

The middle panel shows the ratio of the total grain emission to the total luminosity in the incident continuum. At low *U* the grains absorb very little of the ionizing radiation and the continuum they produce is weak relative to the incident stellar continuum. At large *U* the nebulae are increasingly dust-bounded with nearly the entire incident continuum absorbed by grains. In these cases the grains serve as a bolometer and directly measure the intensity of the incident continuum.

The lowest panel shows the ratio of the Hβ intensity predicted by the models relative to the intensity expected from equation 2. For low *U* the ratio is near unity, showing that most ionizing radiation is absorbed by hydrogen and that hydrogen recombination lines provide an accurate measure of the ionizing continuum. As *U* increases and the nebulae become more dust-bounded this measure fails by increasing extents.

The hydrogen lines are stronger than expected for pure recombination in the low stellar temperature – log *U* corner. The parameters for this grid are so extreme that the clouds in this region are almost totally neutral. The hydrogen emission is enhanced by direct fluorescent excitation of the hydrogen atom by the ~1000Å continuum in these simulations.

## 2.3. A variety of grids

This discussion and the results in Figure 2 suggest a way to measure the luminosity of the continuum source. At low *U* the hydrogen lines count the ionizing photons, and at high *U* the grains act as a bolometer. The ratio IR/Hβ can serve as a *U* indicator.

In this section we examine far larger ranges of astrophysical environments to see whether the results for the low-density Orion simulations are a universal property of nebulae. We computed a very large number of grids ranging to extreme cases. The results are presented in Figure 3, which contain data for nearly $10^3$ model nebulae. The legend for this figure indicates the various sets of results. Each is described below, with the legend title introducing the paragraph.

**BB HII**: This is illuminated by black bodies with temperatures ranging from 25,000 K to 50,000K. The Orion mix was used, log U varied from –4 to 1, and the hydrogen density is $10^2$ cm$^{-3}$.

**OriN2**: This is the grid of results described in the previous section. The abundances and grains will be referred to as the Orion mix.

**OriN4**: This is the same as OriN2 but with a hydrogen density of $10^4$ cm$^{-3}$.

**EnOr2**: Kurucz atmospheres with the temperature ranging from 30,000 to 50,000 K are used, and log U varied from –4 to 0.5. The density was n = $10^2$ cm$^{-3}$. The metallicity of the ISM increases towards the center of the galaxy (Pagel 1997). This model is motivated by HII regions found in inner parts of spiral galaxies: both the metal and grains abundances are increased by a factor of two. In the following this will be referred to as the enhanced mixture.

**EnOr4**: Same as EnOr2 but with a density of $10^4$ cm$^{-3}$.

**PNn2**: This has the Orion mix but a temperature range appropriate for planetary nebulae. The ionizing continua were black bodies with temperatures ranging from 50,000 to 200,000K, log U from –4 to 1, and a density of $10^2$ cm$^{-3}$.

**PNn4**: The same as same as PNn2 but with a density of $10^4$ cm$^{-3}$.

**PNab**: The same as PNn4 but with PNe abundances and grains (Aller & Czyzak 1983; Volk 1991).

**Crab**: The observed continuum of the Crab nebula is used (Davidson & Fesen 1985). This is an example of the hardest possible continuum shape, since the Crab pulsar is an impressive gamma ray source. The enhanced abundance and grain mix was used. Log U varied from –4 to –1, and the hydrogen density varied from $10^2$ to $10^7$ cm$^{-3}$.

**NLR**: This used the Orion mix, a continuum shape representative of a typical Active Galactic Nucleus (AGN; Mathews and Ferland 1987), a density range of $10^2$ to $10^8$ cm$^{-3}$, and log U from –4 to -1.

**NLR2**: This is the same as NLR but with grains and metals increased by a factor of two.

**BLR**: This is the same as NLR except that the density varied from $\log(n_H)$ = 8.5 to 11, to be appropriate for the broad line regions (BLR) of AGN. Log U was varied between –4 and -1. This is offered as the most extreme possible set of models. Densities are high enough for many line transfer and collisional effects to be important. The physical conditions in a BLR cloud are vastly different from those found in classical nebulae (Netzer 1990). Note that the grains are well above their sublimation temperature for these conditions. This could happen in nature if the BLR is a dynamic outflowing wind, since the grains they could survive for the short time they are exposed to the continuum (Bottorff et al. 1997).

## 3. Discussion

Figure 3 suggests a method of measuring either the number of ionizing photons emitted, or the source's bolometric luminosity. The nebulae are parameterized by the ratio of the total thermal IR emission to any hydrogen emission line. (This ratio is the x-axis for the Figure 3.) Hβ is used as a standard in this paper, but the ratio can be converted to any other hydrogen line by using the numerical tables of Storey & Hummer (1995). For instance, an IR/Hβ ratio of 100 is equivalent to an IR/Brγ ratio of 3570, or a IR/Pα ratio of 294.

This analysis will only be valid if both the entire hydrogen recombination line and infrared continuum sources are included in the beam. It is possible to imagine geometries where the regions could be separated by significant amounts, especially for softer continuum sources.

Two limits are clearly present. In the first, when IR/Hβ < 100, grains have little effect. Standard hydrogen photon counting arguments (equation 2) are valid, and the hydrogen lines provide a direct probe of the luminosity in the incident continuum. From the figure it appears that equation 2 is valid to within 20%.

The second limit is the dust-bounded case, where IR/Hβ $\gg$ 100. The incident continuum is mostly absorbed by grains, which act as a bolometer. In this case the IR luminosity nearly represents the total. For the range of models examined, the relationship $L_{tot} = 0.8\ L_{IR}$ is accurate to 20%.

The intermediate case, $10^2 \le$ IR/Hβ $\le 10^3$, is more problematic. No model independent indicator is present. The best strategy would be to use the larger of the two indicators, and take these as a lower limit. Better estimates could be made with if other spectroscopic data were available constrain the models.

We specifically do not analyze any particular object in this paper. Aperture effect will play a role, since thermal IR detectors and those used to measure IR hydrogen lines are generally quite different. However, given a homogeneous data set and the results we have outlined here, luminosities can be measured with relatively limited data set. Finally, an electronic form of the numerical results shown in Figure 3 can be obtained from the authors upon request.

We thank the referee for a thoughtful and complete review of the underlying assumptions in this paper. Research in Nebular Astrophysics at the University of Kentucky is supported by the NSF through grant AST 96-17083.

# 4. Figures

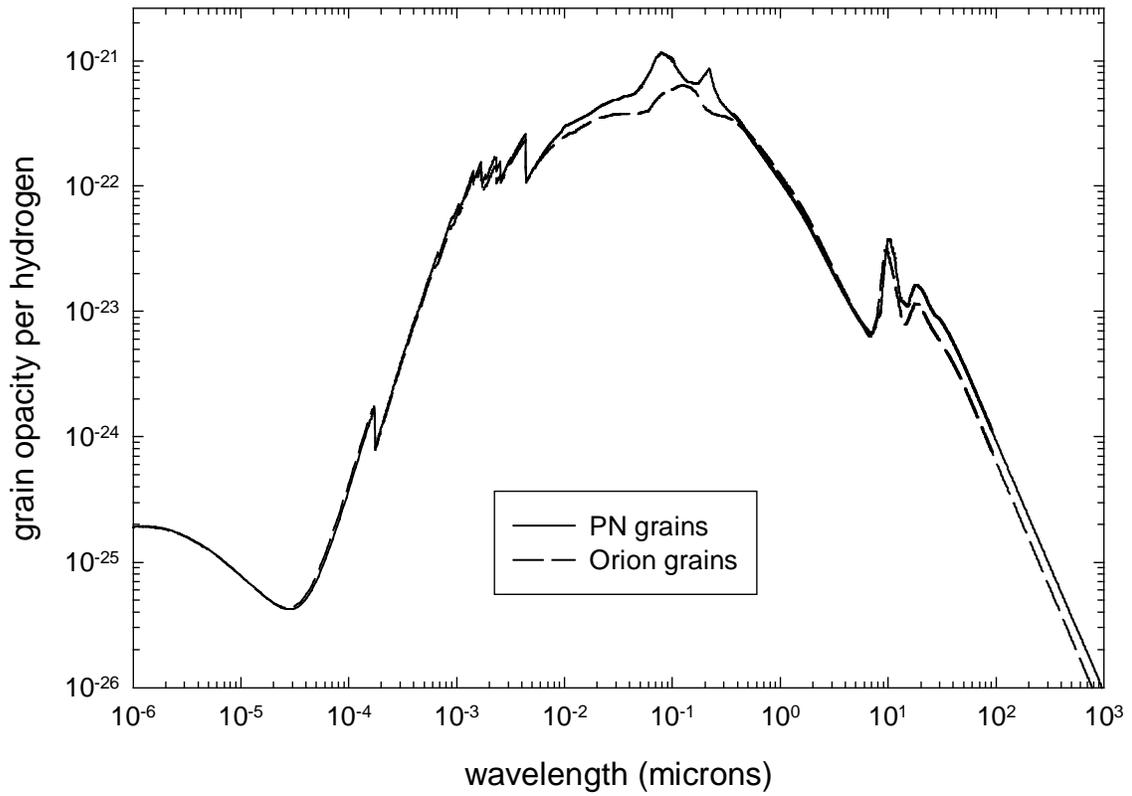

**Figure 1** The absorption opacity (cm$^{-1}$) per hydrogen for the standard mixtures of the two types of grains used in our calculations.

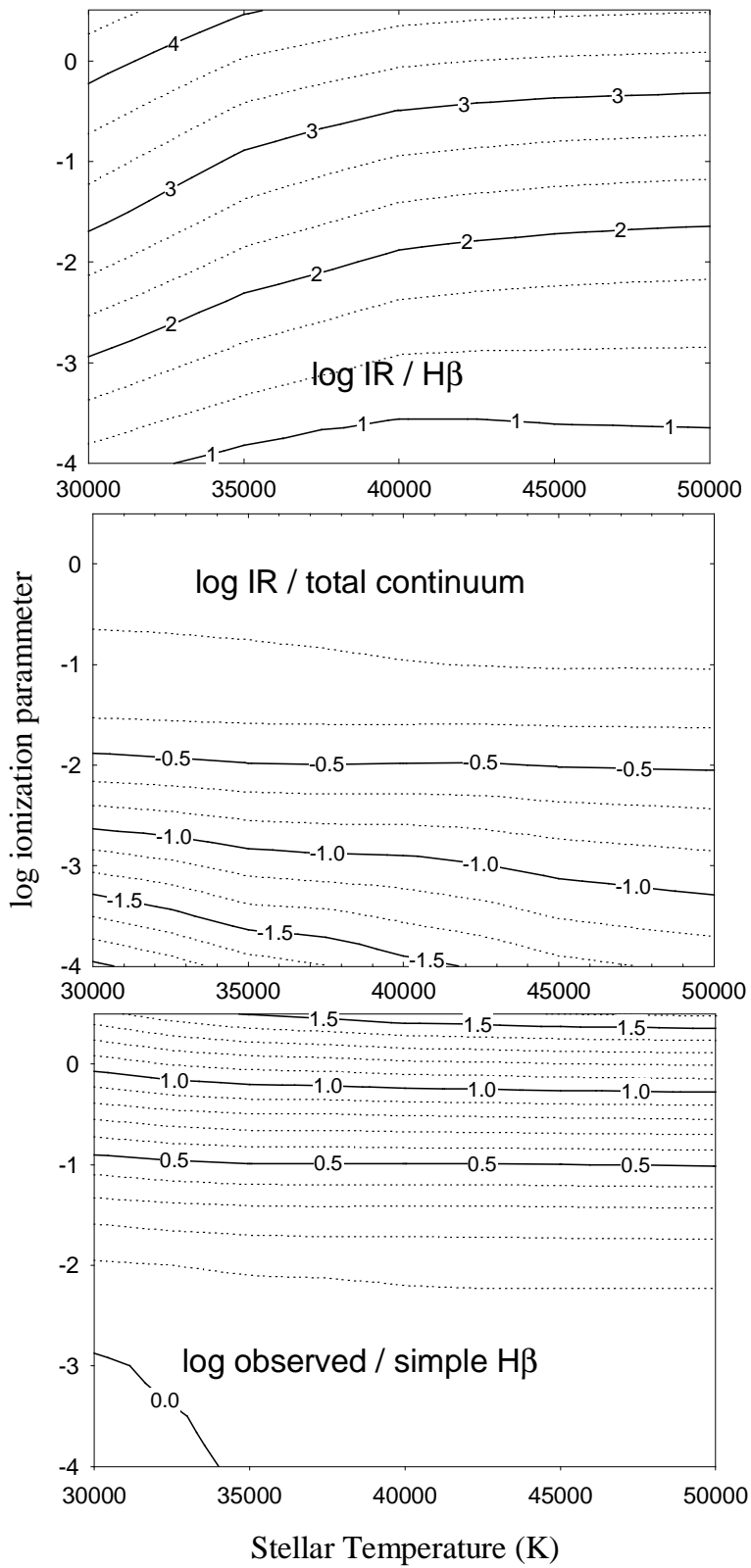

**Figure 2** Results for the HII calculations with Orion abundances and a variety of Kurucz continua. The quantities are described in the text.

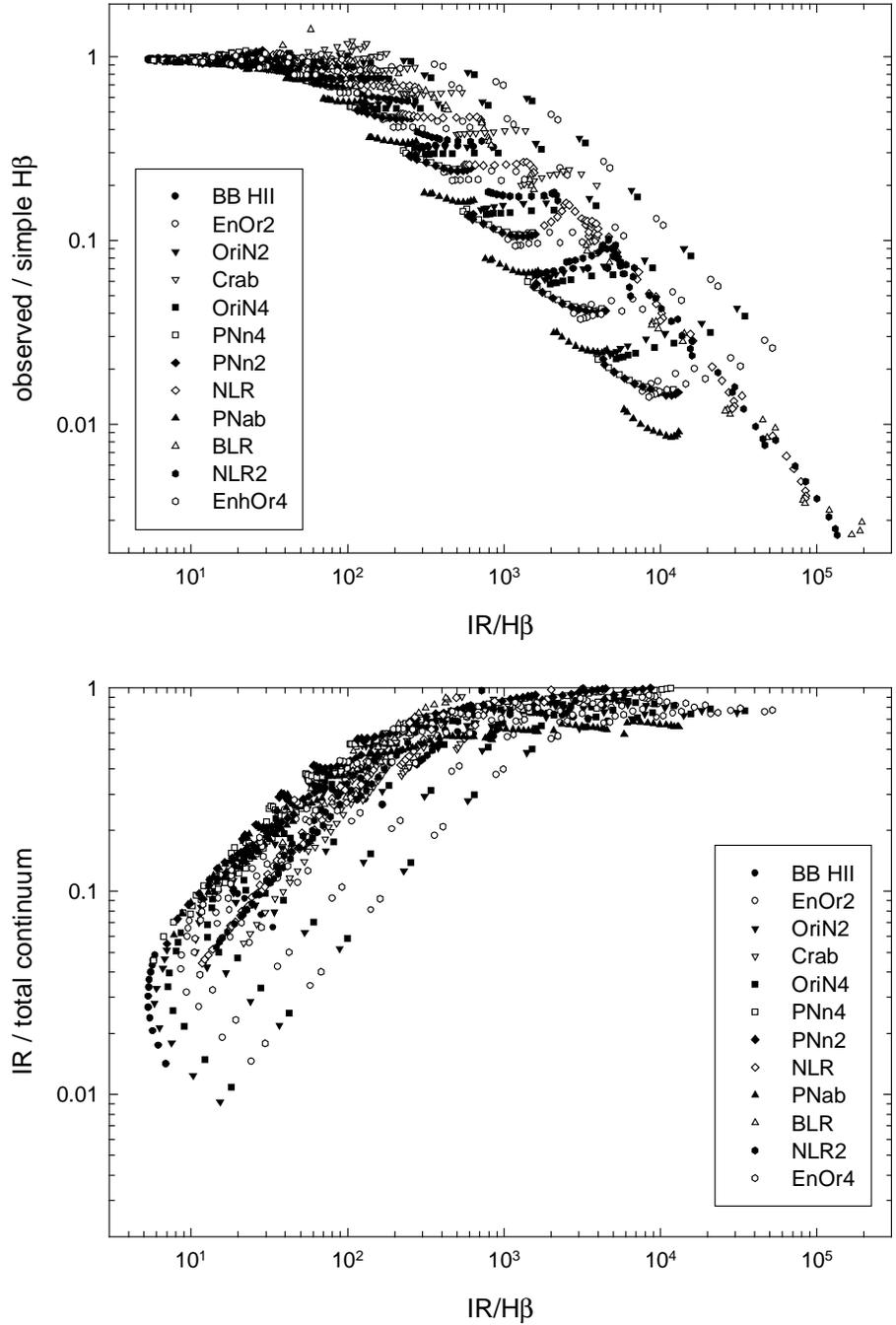

**Figure 3** This summarizes the results of all of our calculations. The x-axis is the ratio of the thermal IR to Hβ, and this serves as an indicator of the relative importance of grains and hydrogen in absorbing the incident continuum. The upper panel shows the ratio of the observed to expected intensity of Hβ. Hydrogen lines count the number of ionizing photons when IR / Hβ < 100. The lower panel shows the thermal IR relative to the total intensity of the incident continuum. The grains act as a bolometer when IR / Hβ >10$^3$.